# SVC-Probe: A Framework for Evaluating Perturbation Generalization in Spatial Foundation-Model Embeddings

Jake Y. Chen[1,2,*], Huu Phong Nguyen[1], Fuad Al Abir[1], Ehsan Saghapour[2]

1 Department of Biomedical Informatics and Data Science, University of Alabama at Birmingham, Birmingham, AL, USA.

2 System Pharmacology and AI Research Center, University of Alabama at Birmingham, Birmingham, AL, USA.

*corresponding author: jakechen@uab.edu

**Abstract (183 / 250 words).**

This work examines perturbation generalization in spatial foundation-model embeddings derived from fluorescence microscopy images. Although these models can discriminate drug conditions accurately, it remains unclear whether the learned representations reflect patterns consistent with expected perturbation axes that transfer across drugs. We introduce SVC-Probe, a perturbation-aware framework that combines Subcellular Embedding Atlas Stability, Mondrian Neighborhood Graphs, and a Foundation Model Perturbation Probe to assess embedding stability, neighborhood rewiring, and centroid prediction under drug treatment. Applied to the CM4AI MDA-MB-468 chemical-perturbation atlas comprising 462 antibody labels and SubCell 1536-dimensional embeddings, SVC-Probe demonstrates that 98.6% three-way condition accuracy does not correlate with reliable cross-drug prediction, with cosine similarity diminishing from 0.944 in-domain to 0.30 under leave-one-drug-out evaluation, constituting a two-drug stress test rather than a general benchmark. Null calibration indicates that raw residual–turnover coupling is largely influenced by generic embedding structure, whereas a drug-specific signal emerges under vorinostat and is consistent with chromatin-related reorganization. In contrast, the paclitaxel axis is not robustly reconstructed, likely due to sparse coverage of microtubule-associated proteins. Together, these results introduce and demonstrate a reusable diagnostic framework for stress-testing spatial virtual-cell representations and indicate that perturbation generalization may serve as a stricter and more informative benchmark than baseline condition discrimination.

## 1 Introduction

Advances in spatial proteomics and microscopy foundation models have enabled the acquisition of concise representations of protein organization from large-scale cellular imaging data [1–3]. These advancements have enhanced the prospect of evaluating spatial virtual-cell representations: computational models that integrate cellular architecture, protein localization, and disease-associated remodeling across scale. Nevertheless, most contemporary spatial foundation models are predominantly assessed on baseline localization, retrieval, or condition discrimination, rather than on their ability to recover perturbation-induced reorganization [2, 3].

In neighboring single-cell domains, perturbation response has emerged as a key standard for representation learning. Models such as scGen, CellOT, and GEARS are evaluated according to whether they can predict out-of-sample responses to chemical or genetic perturbations, rather than simply separate control and treated states [4–6]. This distinction is crucial as an effective representation should preserve mechanistic structure under intervention. In spatial proteomics, the same requirement is arguably more stringent, since treatment effects may manifest not only as protein-level displacement but also as changes in local neighborhood organization and coordinated subcellular remodeling.

A further challenge is biological interpretation. Spatial neighborhoods are not merely geometric entities; they frequently reflect coordinated functional states within tissues and cellular systems [7]. Consequently, a perturbation-aware analysis of spatial embeddings should address at least three interconnected questions: which proteins exhibit stability or drift after treatment, how local neighborhoods reorganize, and whether pretrained latent representations can predict mechanistically meaningful remodeling across drugs. These questions are especially relevant in the present context, as paclitaxel and vorinostat offer complementary perturbation axes centered on microtubule dynamics and chromatin regulation [8, 9].

To address this gap, we introduce SVC-Probe, a framework for assessing perturbation generalization in spatial foundation-model embeddings. The framework integrates three interrelated components: Subcellular Embedding Atlas Stability for protein-level embedding analysis, Mondrian Neighborhood Graphs for neighborhood-level reorganization, and a Foundation Model Perturbation Probe for cross-drug prediction. Its role is to integrate protein stability, neighborhood rewiring, and cross-perturbation prediction within a single evaluation workflow on a common spatial substrate. We apply SVC-Probe to a spatial proteomic perturbation atlas of CM4AI MDA-MB-468 cells under control, paclitaxel, and vorinostat conditions, and use the resulting analyses to shift evaluation from static representation quality toward perturbation generalization. Rather than asking only whether embeddings can distinguish experimental conditions, the proposed framework asks whether they preserve biologically informative structure under intervention, thereby enhancing the credibility of spatial virtual-cell modeling.

## 2 Data and Methods

### *2.1 Background and Related Work*

CM4AI was established under the NIH Bridge2AI program as the Functional Genomics Grand Challenge [10]. The dataset captures 462 antibody-labeled proteins in MDA-MB-468 triple-negative breast cancer cells under three conditions: untreated control, paclitaxel (a microtubule stabilizer [8]), and vorinostat (a pan-HDAC inhibitor [9]). The two drugs were deliberately chosen as they operate on orthogonal axes; a method recovering the same proteins under both drugs is most likely capturing a generic embedding artifact rather than drug-specific mechanism. CM4AI supplies pre-computed DenseNet-121 [11] embeddings alongside the images, which we use as the baseline, in addition to the SubCell pre-trained model [2] for all downstream modules. DenseNet-121 and ConvNeXt-tiny serve only as condition-classification comparators; all SEAS, MNG, and FMP analyses use SubCell embeddings exclusively. These two tasks are categorically distinct: condition classification is a discriminative problem based on observed embeddings, whereas FMP assesses whether the model can predict an unseen drug-condition centroid, relying solely on the control centroid and a drug indicator.

### *2.2 Data and Preprocessing*

We utilized the CM4AI June 2025 release (DOI 10.18130/V3/F3TD5R), which comprises 463 proteins imaged in MDA-MB-468 cells at a resolution of 1536×1536 pixels across four channels (DAPI, target, tubulin, calreticulin), with approximately 8 fields of view (FOVs) per protein-condition pair, yielding a median of approximately 9 cells per protein-condition after Cellpose segmentation. One label was excluded due to incomplete bundled embeddings, resulting in N = 462 proteins with all three conditions. Per-cell SubCell embeddings (1536-D) were generated using the publicly available SubCell checkpoint hosted by Chan Zuckerberg Biohub SF [2], without additional fine-tuning.

Before applying SVC-Probe to the full panel, we established a 33-label calibration subset (15 chromatin modifiers, 15 metabolic enzymes, 3 structural controls: vimentin, pan-actinin, pan-tubulin). This subset was established prior to data access and utilized exclusively for calibration. The two drug arms are intentionally asymmetric: vorinostat directly targets 15 of the 33 chromatin proteins (sensitivity arm), whereas paclitaxel affects only the single pan-tubulin bundle (specificity arm). The paclitaxel arm primarily serves as a contrasting perturbation for cross-drug stress testing; direct microtubule-axis biological interpretation is underpowered in this panel given sparse coverage of microtubule-associated proteins.

## *2.3 SVC-Probe Methods*

SVC-Probe operationalizes three diagnostic questions as independent but complementary modules (Figure 1): how stable is each protein's embedding under perturbation (SEAS), how do co-localization neighborhoods reorganize (MNG), and can foundation-model latents predict treatment-induced centroid shifts (FMP)? All modules share the same input: $\ell_2$-normalized per-cell SubCell embedding clouds, with cell boundaries defined by Cellpose [15].

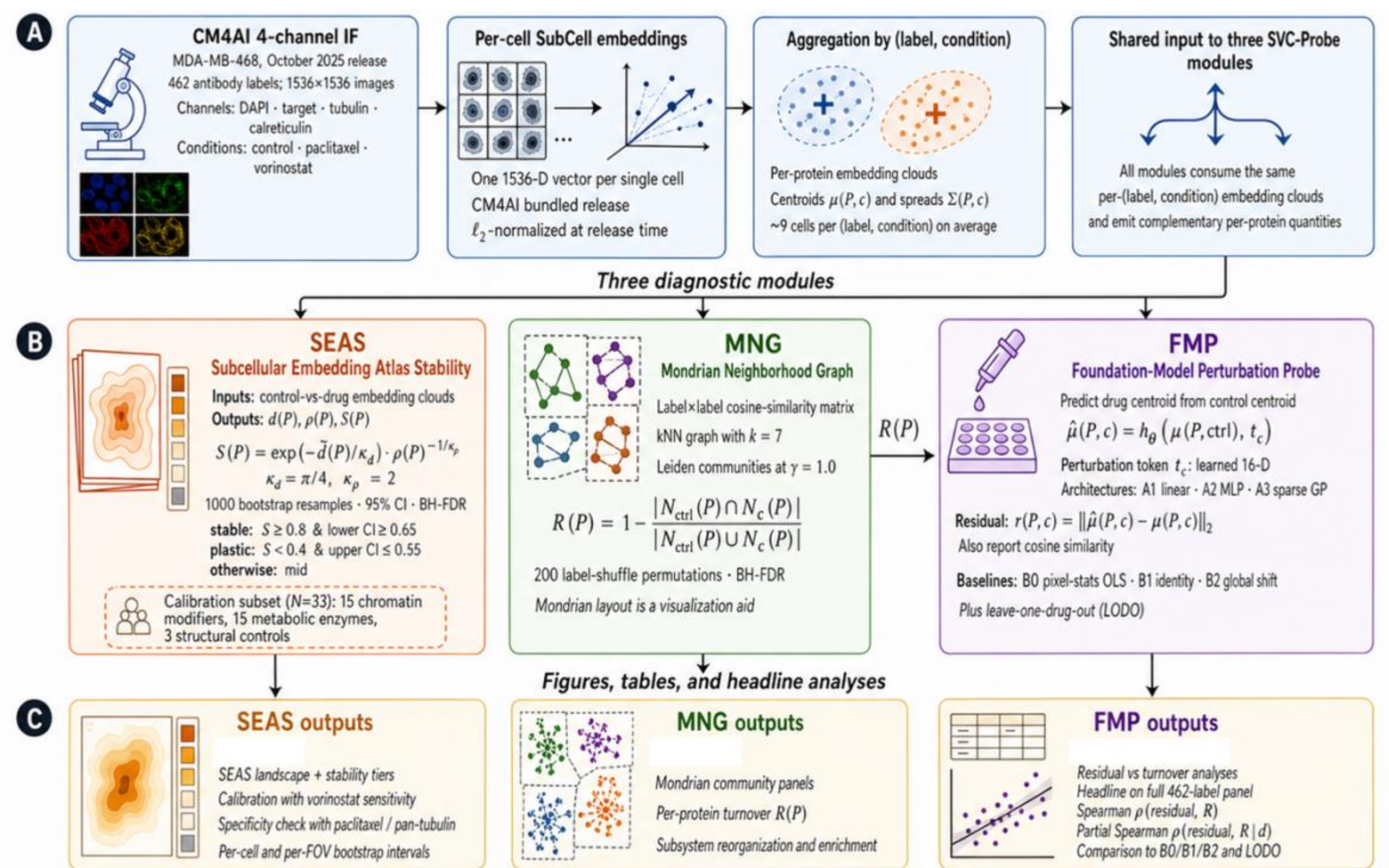

*Figure 1. SVC-Probe pipeline: CM4AI IF → SubCell embeddings (1536-D) → SEAS (stability scoring) → MNG (Mondrian kNN graph + R(P)) → FMP (perturbation probe).*

### 2.3.1 SEAS

In this pipeline, for each protein P and each control-vs-drug condition pair (ctrl, c), SEAS computes $S(P) \in (0, 1]$ capturing how stable the protein's subcellular fingerprint is across the perturbation. With per-cell embedding cloud, centroid $\mu(P,c)$, mean per-dimension variance $\sigma^2(P,c)$, geodesic centroid drift $\tilde{d}(P) = \arccos\langle \hat{\mu}(P,c), \hat{\mu}(P,\mathrm{ctrl})\rangle$, spread ratio $\rho(P) = \max(\sigma^2)/\min(\sigma^2)$.

$$S(P) = \exp(-\tilde{d}(P) / \kappa_d) \cdot \rho(P)^{-1/\kappa_\rho}, \quad \kappa_d = \pi/4, \quad \kappa_\rho = 2$$

Bootstrap CIs: 1000-sample bootstraps over per-cell embeddings give 95% CIs on S(P); BH-FDR-corrected p-values across the panel. Tier setting: stable if $S \geq 0.8$ and lower-CI $\geq 0.65$; plastic if $S < 0.4$ and upper-CI $\leq 0.55$; mid otherwise.

### 2.3.2 MNG

In the second pipeline, per condition c, MNG constructs a label × label cosine-similarity matrix based on protein centroids and preserves the mutual top-k neighbors ($k = 7$; sensitivity over $k \in \{3, 5, 7, 10, 15\}$). Per-protein neighborhood turnover $R(P) = 1 - |N_{ctrl}(P) \cap N_{drug}(P)| / |N_{ctrl}(P) \cup N_{drug}(P)| \in [0, 1]$ captures how much the protein's co-localization neighborhood has reorganized. Leiden community detection ($\gamma = 1.0$) [16] delineates subsystems; each community is laid out as a Mondrian-style axis-aligned rectangle, scaled by node centrality, with the color scheme consistent with the control condition across all three panels.

#### 2.3.3 FMP

For the last pipeline, FMP trains a regression head $h_\theta$ that utilizes the control centroid $\mu(P,ctrl)$ together with a learned 16-dimensional perturbation token to predict the drug-condition centroid $\hat{\mu}(P,c)$. Architectures A1 (linear) and A2 (two-layer multilayer perceptron MLP with hidden 512 units, GELU activation); 5-fold protein-disjoint cross-validation; loss function = MSE + cosine penalty ($\lambda = 1$). Per-protein residual is defined as $r(P,c) = \|\hat{\mu}(P,c) - \mu(P,c)\|_2$ in SubCell space. Baselines: B0, closed-form OLS utilizing 8-dimensional per-channel pixel-statistical summaries (Ahlmann-Eltze-style null [14], intentionally modest); B1, identity (predict drug = control); B2, global drug-mean shift. The LODO variant serves as the primary out-of-distribution assessment: it trains on all drugs except one, then predicts the held-out drug's centroid. This tests whether the learned perturbation token captures a drug-generic transformation or merely overfits the training drug's embedding geometry. Null calibration: $R_{drug\text{-}specific}(P) = R_{real}(P) - \bar{R}_{null}(P)$ (mean of 200 control vs control half-splits). The headline statistic is partial Spearman $\rho(\text{residual}, R \mid d)$, which accounts for raw centroid drift $d(P)$; controlling for drift distinguishes trivial under-prediction of large shifts from the substantive finding that the foundation model overlooks the rewiring axis specifically.

## 3 Results

### 3.1 Classification vs. Perturbation Generalization

SubCell achieved high three-way drug-condition classification accuracy (98.6%), outperforming ConvNeXt-tiny (87.9%) and DenseNet-121 (56.6%), indicating that the embeddings strongly separate observed treatment conditions. The perturbation-prediction task, however, showed a different pattern. In the in-domain setting, the learned A1/A2 perturbation probe achieved the highest cosine similarity (0.944), compared with B0 pixel-statistical null (0.912), B1 identity (0.914), and B2 global shift (0.921). When evaluated under leave-one-drug-out (LODO), the learned model dropped markedly to 0.30. In contrast, B1 remained unchanged from 0.914 to 0.914, as it does not depend on drug-specific fitting, and B2 decreased only slightly from 0.921 to 0.914. B0 was the only baseline to fail substantially under LODO, decreasing from 0.912 to −0.049. These results indicate that the LODO degradation is not simply a consequence of the evaluation protocol. Rather, the learned perturbation probe captures structure that is useful in-domain but does not transfer reliably to an unseen drug, whereas simple control-proximity and global-shift assumptions remain comparatively stable.

| Method | Cosine | Interpretation |
|---|---|---|
| B0 pixel-stat null | 0.912 | Simple image-stat baseline |
| B1 identity | 0.914 | Drug centroid close to control |
| B2 global shift | 0.921 | Global drug shift helps slightly |
| A1/A2 learned (in-domain) | 0.944 | Strong in-domain fit |
| A1/A2 LODO | 0.30 | Poor cross-drug transfer |

Table 1. FMP probe performance. Cosine similarity between predicted and observed drug-condition centroids. B0–B2 are baselines; A1/A2 are the learned perturbation-token models. LODO = leave-one-drug-out.

### 3.2 Null Calibration and Drug-Specific Signal

In this experiment, we asked whether the apparent correlation between FMP residual and neighborhood turnover reflects drug-specific biology or generic embedding structure. To test this, we split the control cells into two pseudo-groups 200 times, computing a same-condition null coupling as a baseline. The null yields $\rho = +0.213$ (95 % CI [+0.138, +0.287]), matching the real drug coupling in magnitude and confirming that raw R(P) is driven by embedding-cloud geometry rather than drug biology. Null calibration is therefore a prerequisite for any drug-specific inference.

Initially, after null subtraction, a drug-specific signal emerges under vorinostat ($\rho = -0.186$, $p = 5.7\times10^{-5}$, N = 462; Cohen's $d \approx 0.44$), whereas paclitaxel shows no consistent pattern ($\rho = -0.043$); the negative sign indicates that proteins with stronger vorinostat-specific neighborhood rewiring tend to have lower FMP residual, consistent with the foundation model encoding a chromatin-axis signal. Next, to assess whether neighborhood turnover contributes information beyond raw centroid displacement, a partial Spearman correlation controlling for drift d(P) yields $\rho = +0.250$ ($p = 7.5\times10^{-8}$), suggesting that R(P) carries information about the rewiring axis that drift alone does not explain; notably, ρ(residual, drift) is itself negative (−0.23), meaning the model is slightly more accurate for high-drift proteins, so drift is not driving the vorinostat signal. Finally, centroid drift and neighborhood turnover are correlated ($\rho \approx 0.43$, $p \ll 10^{-9}$) but not redundant, supporting that SEAS and MNG capture distinct aspects of perturbation response and that running both modules is justified.

### 3.3 SEAS Stability on Calibration Subset

On the 33-label calibration subset, SEAS performs as anticipated: all 15 chromatin-modifier labels are categorized in the plastic or mid tier under vorinostat (KAT2A: S = 0.34 [0.25–0.42]; DNMT3A: S = 0.40 [0.28–0.49]), whereas structural controls remain mid under both drugs. The direction is correct (median S decreases from 0.566 to 0.549 when transitioning from paclitaxel to vorinostat); nevertheless, with N = 33 and around 9 cells per protein-condition, the bootstrap confidence intervals (CIs) are broad (Wilcoxon p = 0.22). Figure 2 illustrates the SEAS landscape.

### 3.4 Chromatin Gene Set Rewiring under Vorinostat

The three pre-specified gene sets with the most robust mechanistic rationale for vorinostat (HDAC family n = 12, HAT writers n = 7, DNA-methylation writers n = 6) exhibit greater turnover under vorinostat compared to paclitaxel (paired Wilcoxon $q_{BH} = 0.078$). This is above the conventional 0.05 threshold ($q_{BH} = 0.078$); therefore hypothesis-generating, and it exhibits leave-one-out instability. The three sets share a unified chromatin-regulatory axis, indicating a correlated signal rather than three independent confirmations. BET readers and histone-methylation writers do not show significant differential turnover; nor do the paclitaxel-mechanism gene sets show significant differential turnover.

Additional FMP metrics, drug-specific turnover analyses, robustness checks, and extended limitations are provided in the Supplementary Material.

## 4 Conclusions

SVC-Probe demonstrates that strong aggregate drug-condition classification does not imply cross-drug perturbation generalization in spatial foundation-model embeddings. Although the evaluated embeddings support highly accurate discrimination of observed drug conditions, their performance degrades substantially under leave-one-drug-out evaluation, indicating limited transfer of perturbation structure across drugs. Raw neighborhood turnover is confounded by generic embedding structure and must be null-calibrated before any drug-specific interpretation. After calibration, directional chromatin-axis signals emerge under vorinostat, consistent with known HDAC inhibition biology, although these results remain hypothesis-generating rather than confirmatory. The pipeline provides a reusable diagnostic workflow for stress-testing spatial virtual-cell representations under perturbation, and highlights that perturbation generalization is a stricter and more informative benchmark than baseline condition discrimination.

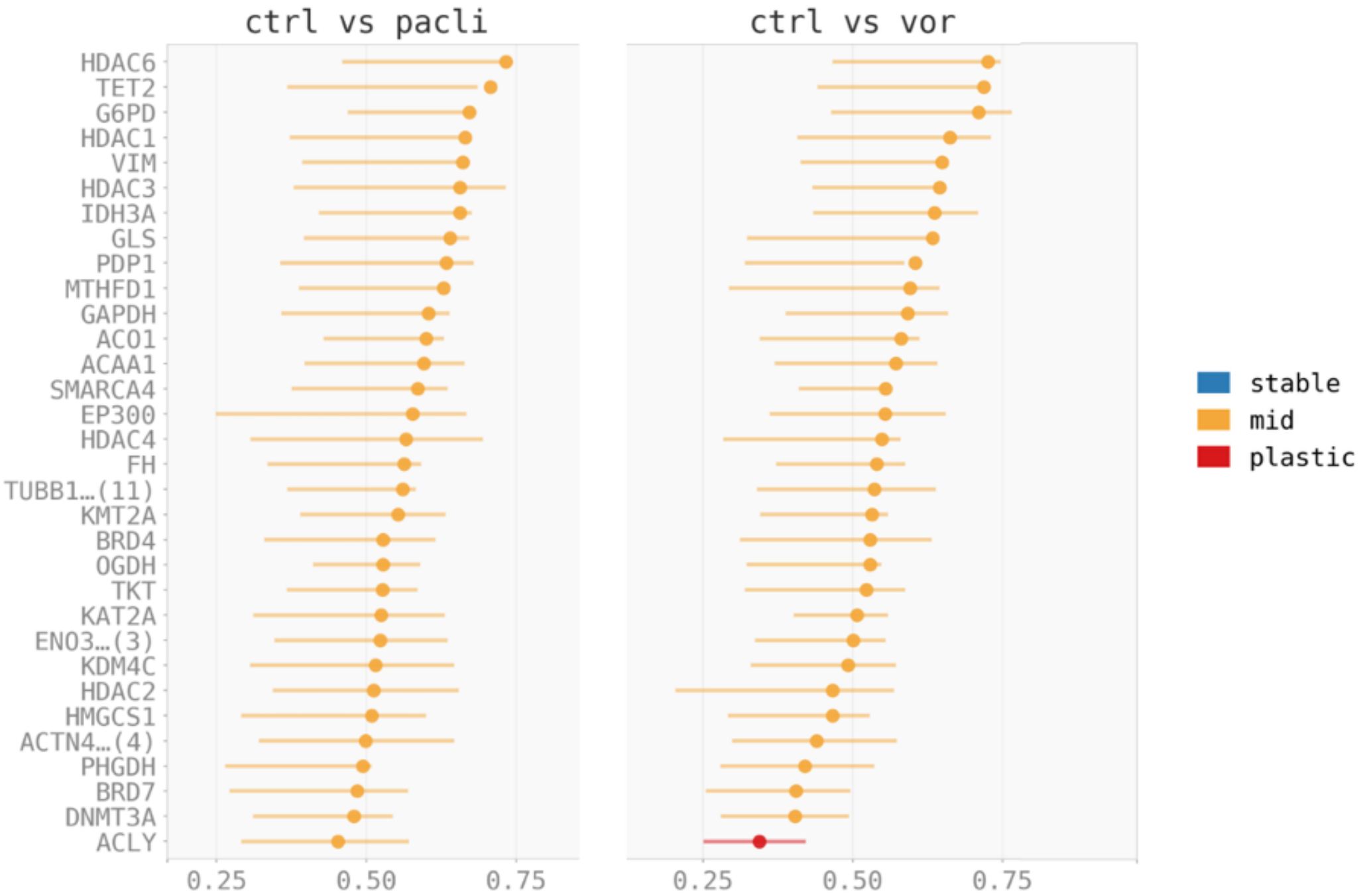

*Figure 2. SEAS stability scores for the 33-label calibration subset under control-versus-paclitaxel and control-versus-vorinostat comparisons. Each point represents the SEAS score of one protein, and horizontal bars indicate 95% bootstrap confidence intervals estimated from per-cell SubCell embeddings. Proteins are shown in the same order across both panels to facilitate comparison between perturbations. Colors indicate SEAS stability tiers, with lower scores corresponding to greater perturbation-associated embedding instability and higher scores corresponding to more stable subcellular embedding profiles.*

Future work should evaluate SVC-Probe across additional drug classes, cell lines, tissue contexts, and imaging protocols to determine whether the observed gap between condition discrimination and perturbation generalization extends beyond the current two-drug CM4AI setting. The framework may also support mechanism-oriented drug profiling, prioritization of perturbation-sensitive proteins, and development of representation-learning objectives that better preserve transferable perturbation structure.

## Conflict of interests

The authors declare no conflict of interest.

## Funding

This work was supported by the NIH Common Fund Data Ecosystem (CFDE) program under award U54OD036472, and by SPARC, UAB.

## Availability of supplementary and source code

https://doi.org/10.5281/zenodo.20897840

## References

[1] Thul et al., “A subcellular map of the human proteome,” Science, vol. 356, p. eaal3321, 2017.

[2] Gupta et al., “SubCell: Proteome-aware vision foundation models for microscopy capture single-cell biology,” bioRxiv 2024.12.06.627299, 2024.

[3] Shaban et al., “A foundation model for spatial proteomics,” arXiv:2506.03373, 2025.

[4] Lotfollahi et al., “scGen predicts single-cell perturbation responses,” Nat. Methods, vol. 16, pp. 715–721, 2019.

[5] Bunne et al., “Learning single-cell perturbation responses using neural optimal transport,” Nat. Methods, vol. 20, pp. 1759–1768, 2023.

[6] Roohani et al., “Predicting transcriptional outcomes of novel multigene perturbations with GEARS,” Nat. Biotechnol., vol. 42, pp. 927–935, 2024.

[7] Schürch et al., “Coordinated cellular neighborhoods orchestrate antitumoral immunity at the colorectal cancer invasive front,” Cell, vol. 182, pp. 1341–1359, 2020.

[8] Mitchison, “The proliferation rate paradox in antimitotic chemotherapy,” Mol. Biol. Cell, vol. 23, pp. 1–6, 2012.

[9] West et al., “New and emerging HDAC inhibitors for cancer treatment,” J. Clin. Invest., vol. 124, pp. 30–39, 2014.

[10] Clark et al., “Cell Maps for Artificial Intelligence: AI-ready maps of human cell architecture from disease-relevant cell lines,” bioRxiv 2024.05.21.589311, 2024.

[11] Huang et al., “Densely connected convolutional networks,” in Proc. CVPR, pp. 4700–4708, 2017.

[12] Sun et al., “Generative machine learning unlocks the first proteome-wide image of human cells,” bioRxiv 2026.03.31.715748, 2026.

[13] Mädler et al., “scPortrait integrates single-cell images into multimodal modeling,” bioRxiv 2025.09.22.677590, 2025.

[14] Ahlmann-Eltze et al., “Deep learning-based gene perturbation effect prediction does not yet outperform simple linear methods,” Nat. Methods, vol. 22, pp. 1657–1661, 2025.

[15] Pachitariu et al., “Cellpose 2.0: how to train your own model,” Nat. Methods, vol. 19, pp. 1634–1641, 2022.

[16] Traag et al., “From Louvain to Leiden: guaranteeing well-connected communities,” Sci. Reports, vol. 9, p. 5233, 2019.